\documentclass[twocolumn]{aastex6}

\begin{document}


\title{Discovery of an extremely-luminous dust-obscured galaxy observed with SDSS, WISE, JCMT, and SMA}


\author{Yoshiki Toba 		\altaffilmark{1,2},
		Junko Ueda 			\altaffilmark{3},
		Chen-Fatt Lim  	 	\altaffilmark{1,4},
		Wei-Hao Wang		\altaffilmark{1},
		Tohru Nagao 		\altaffilmark{2},
		Yu-Yen Chang		\altaffilmark{1},
		Toshiki Saito		\altaffilmark{5,6},
		Ryohei Kawabe		\altaffilmark{5,7,8}
		}	
\affil{}  			  
  \altaffiltext{1}{Academia Sinica Institute of Astronomy and Astrophysics, PO Box 23-141, Taipei 10617, Taiwan}
  \email{toba@asiaa.sinica.edu.tw}
  \altaffiltext{2}{Research Center for Space and Cosmic Evolution, Ehime University, 2-5 Bunkyo-cho, Matsuyama, Ehime 790-8577, Japan} 
  \altaffiltext{3}{Harvard-Smithsonian Center for Astrophysics, 60 Garden Street, Cambridge, MA 02138,
USA}
  \altaffiltext{4}{Graduate Institute of Astrophysics, National Taiwan University, PO Box 23-141, Taipei 10617, Taiwan}   
  \altaffiltext{5}{National Astronomical Observatory of Japan, 2-21-1 Osawa, Mitaka, Tokyo, 181-8588, Japan}
  \altaffiltext{6}{Max Planck Institute for Astronomy, K\"onigstuhl 17, D-69117 Heidelberg, Germany} 
  \altaffiltext{7}{Department of Astronomy, The University of Tokyo, 7-3-1 Hongo, Bunkyo-ku, Tokyo 113-0033, Japan}
  \altaffiltext{8}{Department of Astronomical Science, Graduate University for Advanced Studies
(SOKENDAI), 2-21-1 Osawa, Mitaka, Tokyo 181-8588, Japan}


\begin{abstract}
We present the discovery of an extremely-luminous dust-obscured galaxy (DOG) at $z_{\rm spec}$ = 3.703, WISE J101326.25+611220.1.
This DOG is selected as a candidate of extremely-luminous infrared (IR) galaxies based on the photometry from the Sloan Digital Sky Survey and {\it Wide-field Infrared Survey Explorer}.
In order to derive its accurate IR luminosity, we perform follow-up observations at 450 and 850 $\micron$ using the Submillimetre Common User Bolometer Array 2 on the James Clerk Maxwell Telescope, and at 870 and 1300 $\micron$ using the Submillimeter Array, which enable us to pin down its IR Spectral Energy Distribution (SED).
We perform SED fitting using 14 photometric data (0.4 - 1300 $\micron$) and estimate its IR luminosity, $L_{\rm IR}$ (8-1000 $\micron$), to be $2.2^{+1.5}_{-1.0}$ $\times 10^{14}$ $L_{\sun}$, making it one of the most luminous IR galaxies in the Universe.
The energy contribution from an active galactic nucleus (AGN) to the IR luminosity is $94^{+6}_{-20}$\%, which indicates it is an AGN-dominated DOG.
On the other hand, its stellar mass ($M_*$) and star formation rate (SFR) are $\log \,(M_\ast/M_{\sun})$ = $11.2^{+0.6}_{-0.2}$ and $\log \,({\rm SFR}/M_{\sun}\,{\rm yr}^{-1}$) = $3.1^{+0.2}_{-0.1}$, respectively, which means that this DOG can be considered as a starburst galaxy in $M_*$--SFR plane.
This extremely-luminous DOG shows significant AGN and star forming activity that provides us an important laboratory to probe the maximum phase of the co-evolution of galaxies and supermassive black holes.
\end{abstract}




\section{Introduction} 
\label{Intro}

Galaxies with infrared (IR) luminosity, $L_{\rm IR}$ (8-1000 $\micron$), exceeding 10$^{12}$ $L_{\sun}$ shed important light on how galaxies form and evolve throughout the history of the Universe. 
Their IR luminosity is generated by significant star formation (SF) and/or active galactic nucleus (AGN) activity behind a large amount of dust.
The strong ultraviolet (UV) and optical radiation due to the SF/AGN activity is absorbed by the surrounding dust, which then re-emits the enormous energy at the IR wavelength.
Thanks to the advent of IR satellites such as {\it Infrared Astronomical Satellite} \citep[{\it IRAS}: ][]{Neugebauer}, {\it Spitzer Space Telescope} \citep{Werner}, {\it AKARI} \citep{Murakami}, {\it Wide-field Infrared Survey Explorer} \citep[{\it WISE}: ][]{Wright}, and {\it Herschel Space Observatory} \citep{Pilbratt}, extreme galaxies with $L_{\rm IR}$ $>$ 10$^{13}$ $L_{\sun}$ and 10$^{14}$ $L_{\sun}$ (hyper-luminous IR galaxies (HyLIRGs) and extremely-luminous infrared galaxies (ELIRGs), respectively), have been discovered \citep[e.g.,][]{Rowan-Robinson,Tsai}.
Recently, a galaxy (WISE J224612.07-714401.2, hereafter WISE2246) with $L_{\rm IR} = 2.2 \times 10^{14}$ $L_{\sun}$ was discovered \citep{Tsai} and found to be one of the most luminous galaxies with multi-wavelength data in the Universe \footnote{Based on the optical multi-color selection, BR 1202--0725 with $L_{\rm FIR}$ (1-200 $\micron$) = $3.7 \times 10^{14}$ $L_{\sun}$ at $z$ = 4.69 is also known as one of the most luminous galaxies  with multi-wavelength data \citep{Leipski}.}.
Its extreme IR luminosity could indicate that it corresponds to the peak of AGN and/or SF activity, providing the laboratory for understanding the galaxy formation and evolution and connection to their supermassive black holes (SMBHs) under an extreme condition \citep{Hopkins,Ricci}.
However, the volume densities of HyLIRGs/ELIRGs are extremely low, and thus wide and deep surveys are required to detect these spatially rare populations.

For discovering extremely-luminous IR objects and investigating their physical properties, \cite{Toba_16} have performed a systematic HyLIRGs/ELIRGs survey with the Sloan Digital Sky Survey (SDSS) Data Release 12 \citep[DR12:][]{York,Alam} and the ALLWISE catalogs \citep{Cutri}.
They first selected 67 objects with $i$ - [22] $>$ 7.0, where $i$ and [22] are $i$-band and 22 $\micron$ AB magnitude, respectively.
This color selection is used for IR-bright dust-obscured galaxies (DOGs\footnote{The original definition of DOGs was flux density at 24 $\micron$ $>$ 0.3 mJy and $R - [24] > 14$ (corresponding to $F_{\rm 24}/F_{\rm R} > 982$), where $R$ and [24] represent Vega magnitudes in the $R$-band and 24 $\micron$, respectively \citep{Dey}. Our DOGs selection criteria is optimized for $i$-band and 22 $\micron$ flux density \cite[see][for more detail]{Toba}.}) \citep[e.g.,][]{Toba,Toba_17a,Toba_17d}.
They also have spectroscopic redshifts obtained from the SDSS.
They then performed Spectral Energy Distribution (SED) fitting for 67 DOGs and estimated their IR luminosities.
Consequently, they successfully discovered 24 HyLIRGs and a candidate of an ELIRG, WISE J101326.25+611220.1 (hereafter, WISE1013\footnote{Given the best-fit SED of WISE1013 (see Figure \ref{SED}), the estimated flux ratio, $F_{\rm 24}/F_{\rm R} = 1005$ that satisfies the original DOG's criteria defined by \cite{Dey}.}).
Its spectroscopic redshift is $z_{\rm spec}$ = 3.703 and the estimated IR luminosity was $L_{\rm IR} = 1.1 \times 10^{15}$ $L_{\sun}$ \citep[see lower-right panel of Figure 6, and bottom panel of Figure 8 in][]{Toba_16}.
However, since they did not have deep rest-frame mid-IR (MIR) and far-IR (FIR) photometry, the derived IR luminosity has a large uncertainty. 
In order to determine the accurate IR luminosity of this candidate of an ``extremely-luminous DOG'', FIR and submillimter data are strongly required.

In this paper, we present follow-up observations of the candidate of an extremely-luminous DOG, WISE1013,  at 450 and 850 $\micron$ using the Submillimetre Common User Bolometer Array 2 \citep[SCUBA-2: ][]{Holland} on the James Clerk Maxwell Telescope (JCMT), and at 870 and 1300 $\micron$ using the Submillimeter Array \citep[SMA: ][]{Ho}.
Since these observing wavelengths correspond to rest-frame FIR for this object, these observations are critical to constrain the IR-SED.
Throughout this paper, the adopted cosmology is a flat universe with $H_0$ = 70 km s$^{-1}$ Mpc$^{-1}$, $\Omega_M$ = 0.3, and $\Omega_{\Lambda}$ = 0.7. Unless otherwise noted, all magnitudes refer to the AB system.

\section{Data and analysis}
\subsection{A candidate of ELIRGs: WISE1013}
\label{WISE1013}
WISE1013, a candidate of ELIRGs is selected from the sample in \cite{Toba_16}.
The basic information and the measured fluxes (see Section \ref{F_obs}) are summarized in Table \ref{Table}.
This object was reported as an extremely red quasar (ERQ) (SDSS J101326.24+611219.7) by \cite{Hamann}.
The authors selected ERQs based on the SDSS and WISE catalogs by adopting a color cut, ($r$ - [22])$_{\rm vega}$ $>$ 14, which is similar to our selection, and conducted a detail analysis for their spectra.
They reported that this object shows an broad C{\,\sc iv} $\lambda$1549 emission line with a blueshift of  $>$ 2500 km s$^{-1}$.
That unusual spectrum could indicate that ionized gas is outflowing from the ERQ.
Such phenomenon is often observed in red/dust-obscured AGNs \citep[e.g.,][]{Zakamska,Toba_17c}. 

\begin{table}[h]
\renewcommand{\thetable}{\arabic{table}}
\centering
\caption{Observed properties of WISE1013.}
\label{Table}
\begin{tabular}{lr}
\tablewidth{0pt}
\hline
\hline
WISE J101326.25+611220.1		&								\\
\hline
R.A. (SDSS) [J2000.0] 			& 	10:13:26.24					\\
Decl. (SDSS) [J2000.0]			& 	+61:12:19.76  				\\
Redshift (SDSS)					&	3.703 $\pm$ 0.001			\\
SDSS $u$-band [$\mu$Jy]	&	1.26\tablenotemark{a} $\pm$ 0.87	\\
SDSS $g$-band [$\mu$Jy]	&	3.47 $\pm$ 0.47		\\
SDSS $r$-band [$\mu$Jy]	&	13.70 $\pm$ 0.67	\\
SDSS $i$-band [$\mu$Jy]	&	13.58 $\pm$ 0.95	\\
SDSS $z$-band [$\mu$Jy]	&	21.09 $\pm$ 4.03	\\
{\it WISE} 3.4 $\micron$ [mJy]	&	0.05 $\pm$ 0.01				\\
{\it WISE} 4.6 $\micron$ [mJy]	&	0.13 $\pm$ 0.01				\\
{\it WISE} 12  $\micron$ [mJy]	&	3.30 $\pm$ 0.16				\\
{\it WISE} 22  $\micron$ [mJy]	&	10.70 $\pm$ 0.98			\\
{\it AKARI} 90  $\micron$ [mJy]	&	$<$ 0.33\tablenotemark{b}	\\
SCUBA-2 450 $\micron$ [mJy]		&	46.00 $\pm$ 8.05\tablenotemark{c}			\\
SCUBA-2 850 $\micron$ [mJy]		&	13.35 $\pm$ 0.67\tablenotemark{c}			\\
SMA 870 $\micron$ [mJy]			&	13.60 $\pm$ 2.72\tablenotemark{c}	 		\\
SMA 1.3 mm [mJy]				&	6.49 $\pm$ 1.30\tablenotemark{c}				\\
$L_{\rm IR}$ (8-1000 $\micron$) [$L_{\sun}$]			&	$2.2^{+1.5}_{-1.0}$ $\times 10^{14}$	\\
$L^{\rm AGN}_{\rm IR}$ (8-1000 $\micron$) [$L_{\sun}$]	& 	$2.0^{+1.5}_{-1.0}$ $\times 10^{14}$	\\
$L^{\rm SF}_{\rm IR}$ (8-1000 $\micron$) [$L_{\sun}$]	& 	$1.2^{+0.6}_{-0.5}$ $\times 10^{13}$	\\
$\log M_{\rm *}$ [$M_{\sun}$]							& 	$11.2^{+0.6}_{-0.2}$				\\ 
$\log {\rm SFR}$ [$M_{\sun}$/yr]		 				& 	$3.1^{+0.2}_{-0.1}$					\\ 
\hline
\multicolumn{2}{l}{(a) it was used for upper limit (see Section \ref{SED_fit}).}\\
\multicolumn{2}{l}{(b) 3$\sigma$ upper limit.} \\
\multicolumn{2}{l}{(c) it includes both systematic error and RMS noise.}
\end{tabular}
\end{table}

\subsection{Follow-up observations with SCUBA-2 and SMA}
\label{F_obs}
Imaging at 450 and 850 $\micron$ was taken simultaneously with SCUBA-2 on JCMT. 
The observations were conducted under Band-1 condition ($\tau_{\rm 225GHz} <$ 0.05) on 2017 May 8 and 18, as an urgent program (S17AP002, PI: Y.Toba). 
We observed WISE1013 in four 30-minute scans using the compact ``Daisy'' scan pattern. 
The total on-source time is about 2 hours. 
During the observations, we observed the nearby radio source IRC+10216 for pointing check. 
The pointing offsets are typically about 1\arcsec.
All data were reduced using the Sub-Millimeter Common User Reduction Facility \citep[SMURF: ][]{Chapin} and the Pipeline for Combining and Analyzing Reduced Data \citep[PICARD: ][]{Jenness}. 
We adopted the standard ``blank field'' configuration, which is optimized for faint point sources. 
To obtain flux calibration, we observed six calibrators on the same nights under Band-1 weather. 
The averaged Flux Conversion Factors are 503 $\pm$ 88 and 519 $\pm$ 26 Jy beam$^{-1}$ pW$^{-1}$ for 450 $\micron$ and 850 $\micron$, respectively, and are consistent with the nominal values of 491 and 537 Jy beam$^{-1}$ pW$^{-1}$ \citep{Dempsey}.
To optimize the detection, we convolved the maps with broad Gaussian kernels (FWHM of 20$\arcsec$ and 30$\arcsec$ for 450 and 850 $\micron$, respectively) and subtracted the convolved maps from the original maps to remove any large structure. 
Then, we convolved matched-filters to the maps, using Gaussian kernels matched to the instrumental point spread functions (FWHM of 7.5$\arcsec$ and 14$\arcsec$ for 450 and 850 $\micron$, respectively) to obtain optimal signal-to-noise ratio. 
We detect the source at both wavebands. 

Observations at 870~$\mu$m (345~GHz) and 1.3~mm (240~GHz) were carried out with the SMA on 2016 December 28 (2016B-A003, PI: Y.Toba).  
In order to observe at 345~GHz and 240~GHz simultaneously, we used the dual frequency mode of the SWARM correlator, which gave the total bandwidth of 12.6~GHz for each receiver.  
The data were obtained using the compact configuration.  
The FWHM of the primary beam is 32\arcsec at 345~GHz and 46\arcsec at 240~GHz.  
The quasar 3c273 was observed for bandpass calibration,  and the quasar J1048+717 was observed for phase and amplitude calibration.  
Absolute flux calibration was performed using Callisto.  
The uncertainty of flux calibration is 20\%.
Data reduction was carried out using the IDL-based SMA calibration tool MIR, and imaging was done using the MIRIAD package.  
The total observing time on WISE1013 is about 2.3~hours, excluding bad scans.  
The synthesized beam sizes are 2\farcs51 $\times$ 1\farcs67 at 870~$\mu$m  and 3\farcs55 $\times$ 2\farcs33 at 1.3~mm  by adopting natural weighting of the visibilities.  
The achieved RMS noise levels are 2.1~mJy beam$^{-1}$ and 0.7~mJy beam$^{-1}$ in the 870~$\mu$m and 1.3~mm continuum maps, respectively.
   
The flux measurements of the SCUBA-2 and SMA data were performed using the Common Astronomy Software Applications package (CASA\footnote{\url{https://casa.nrao.edu/}}, ver. 4.7.2).
We performed a 2D Gaussian fit for each image and estimated the total fluxes within $10\arcsec \times 10\arcsec$ apertures for the SCUBA-2 data, and $4\arcsec \times 4\arcsec$ apertures for the SMA data.
The measured fluxes are listed in Table \ref{Table}.

\section{Results and discussions}
\subsection{Multi-wavelength images}
\begin{figure*}
   \centering
   \includegraphics[width=0.9\textwidth]{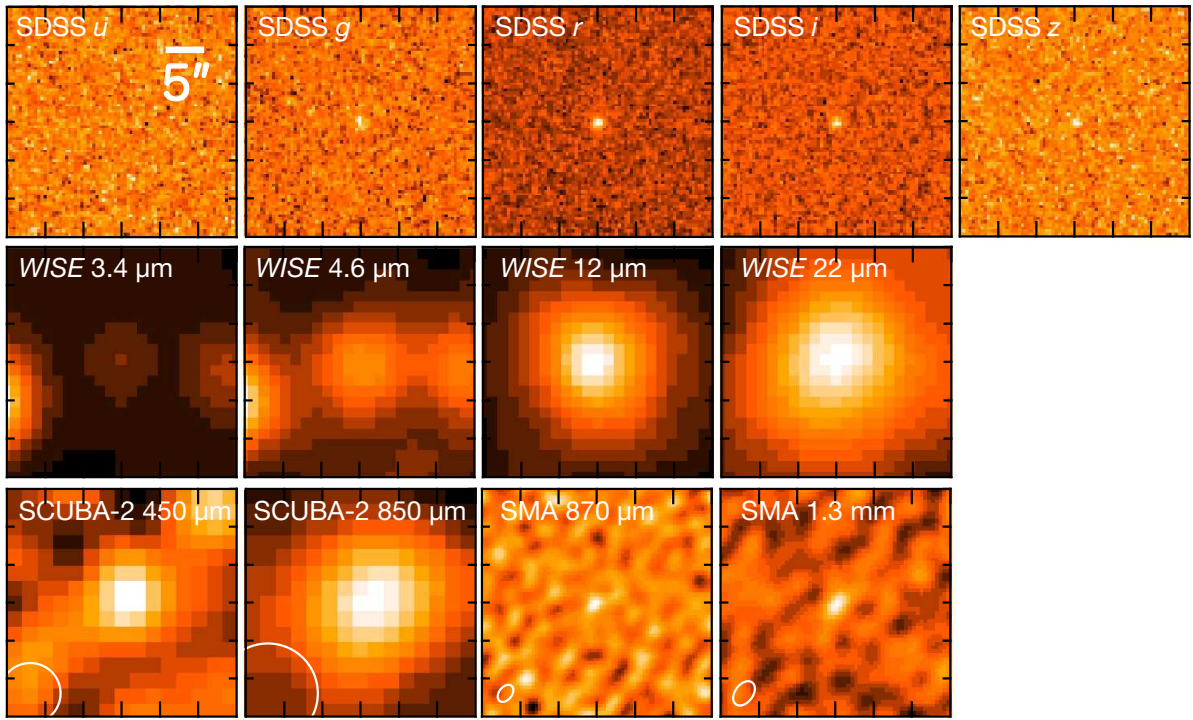}
   \caption{Multi-wavelength images of WISE1013 with a size of $30^{\prime\prime} \times 30^{\prime\prime}$. The white circles on 450, 850, 870, 1300 $\micron$ images are beam sizes for each instrument.}
   \label{image}
   \end{figure*}

Figure \ref{image} shows the multi-wavelength images of WISE1013.
We note that its $u$-band magnitude (23.50 mag) is significantly fainter than magnitude limit (95\% completeness for point source) of the SDSS ($u$-mag = 22.0\footnote{\url{http://www.sdss.org/dr12/scope/}}), and we confirmed that there is no detection in the $u$-band image (Figure \ref{image}).
This source is a complete dropout at $u$-band, because of its high redshift, and thus we used the $u$-band flux density upper limit (see Section \ref{SED_fit}).
We confirm that there is no galaxy within a 30$\arcsec$ radius of WISE1013, as shown in the SDSS images (Figure \ref{image}) and we have SCUBA-2 and SMA images with high angular resolutions, which give us secure estimates of rest-frame FIR fluxes without suffering from flux contamination of neighborhoods.

\subsection{IR luminosity derived from the SED fitting}
\label{SED_fit}

We estimate the IR luminosity of WISE1013 by using the fitting code SEd Analysis using BAyesian Statistics ({\tt SEABASs}: \citealt{Rovilos}).
{\tt SEABASs} produces a best-fit SED by combining three templates (stellar, AGN, and SF components) based on the maximum likelihood method \citep[see ][for more detail]{Rovilos,Toba_17b}.
{\tt SEABASs} gives stellar templates from \cite{Bruzual} stellar population models assuming a \cite{Chabrier} initial mass function (IMF), and each model are reddened by using a \cite{Calzetti} dust extinction law, resulting in the stellar mass and color excess ($E (B-V)$) of a galaxy as an output.
Users can input AGN and SF templates prepared by themselves into {\tt SEABASs} code.

For AGN templates, we input the SED library for \cite{Silva} as the obscured AGN templates, which consists of four torus templates with varying extinction ranging from $N_{\rm H}$ = 0, 10$^{22}$, 10$^{23}$, and  10$^{24}$ cm$^{-2}$.
We also input the library of \cite{Polletta} that provides optically selected AGNs \citep[see ][for more detail]{Polletta}.
In addition to empirical templates, we input the AGN templates\footnote{\url{http://www.eso.org/~rsiebenm/agn_models/}} created by computing a self-consistent three-dimensional radiative transfer code \citep{Siebenmorgen}.
In this model, dust can be considered as a clumpy medium or a homogeneous disk, or as a combination of the two.
We used the SED library of AGN torus models with a set of model parameters; the viewing angle ($\theta$), the inner radius ($r_{\rm in}$), the volume filling factor($\eta$), optical depth of the clouds ($\tau_{V,{\rm cl}}$), and the optical depth of the disk midplane ($\tau_{V,{\rm mid}}$).
For the SF templates, we input the library of \cite{Chary}, \cite{Polletta}, and \cite{Mullaney}, in which we cropped at rest-frame wavelengths below 4.5 $\micron$ to avoid a duplication of the emission from the stellar component in the same manner as \cite{Rovilos}.

We took into account the equilibrium between the energy absorbed from the stellar component and the energy emitted in the IR by the SF.
Although WISE1013 was not detected by the {\it AKARI} FIR all-sky survey \citep{Yamamura}, we used the {\it AKARI} 90 $\micron$ flux 3$\sigma$ upper limit.
Also, we used the SDSS $u$-band flux upper limit.
We performed the SED fitting using 14 photometric points between $u$-band and 1300 $\micron$, and estimated the IR luminosity.

Note that given a limited number of data points, degeneracies between the different fitted components could be occurred when executing the SED fitting with three components.
In order to have an estimate of the uncertainty of the derived quantities, {\tt SEABASs} keeps the likelihood values for every trial fits and estimates the 1$\sigma$, 2$\sigma$ and 3$\sigma$ confidence intervals of a quantity in question by using log-likelihood differences between the trial fits and the best fit \citep[see Appendix A in][for more detail]{Rovilos}.
As a result, {\tt SEABASs} outputs the uncertainties as the 2$\sigma$ confidence interval.
Therefore, the influence of the difference between the inputted SED templates on the derived $L_{\rm IR}$ is included in the uncertainties.
On the other hand, we should keep in mind that since this SED fitting technique tries to find best (combination of) template(s) given the limited numbers of stellar/AGN/SF templates, we do not rule out a possibility of a template that is better able to fit the data.

   \begin{figure*}
   \centering
   \includegraphics[width=0.9\textwidth]{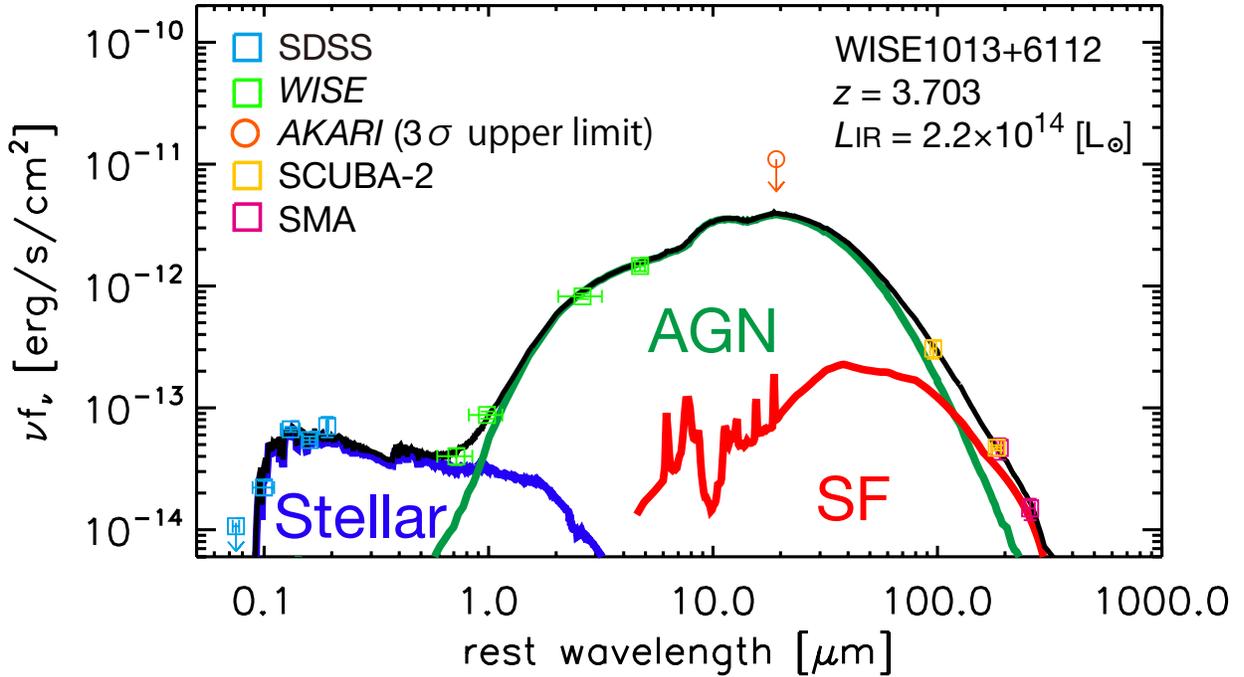}
   \caption{SED of WISE1013. The blue, green, orange, yellow, and pink squares represent the data from SDSS, {\it WISE}, {\it AKARI} (3$\sigma$ upper limit), SCUBA-2, and SMA, respectively. The contribution from the stellar, AGN, and SF components to the total SEDs are shown as blue, green, and red lines, respectively. The black solid line represents the resultant SED. The best-fit stellar template is a template of stellar population with a age of 0.1 Gyr and solar metallicity assuming a $\tau$ model with $\tau$ = 0.3 Gyr in \cite{Bruzual}. The best-fit SF template is ``NGC6090'' \citep{Polletta} with cropping at rest-frame wavelengths below 4.5 $\micron$ while the best-fit AGN template is a template with  $\theta$ = 19\arcdeg, $r_{\rm in}$ = $3\times10^{13}$ cm, $\eta$ = 7.7\%, $\tau_{V,{\rm cl}}$ = 13.5, and $\tau_{V,{\rm mid}}$ = 1000 \citep{Siebenmorgen}.}
   \label{SED}
   \end{figure*}   
   
Figure \ref{SED} shows the resultant SED fitting for WISE1013 from {\tt SEABASs}.
The best-fit stellar template is a template of stellar population with a age of 0.1 Gyr and a metallicity of $Z$ = 0.02 assuming an exponentially declining a star formation history ($\tau$ model) with a timescale (i.e., e-folding time) of $\tau$ = 0.3 Gyr in \cite{Bruzual}.
The best-fit SF template is ``NGC6090\footnote{\url{http://www.iasf-milano.inaf.it/~polletta/templates/swire_templates.html}}'' in \cite{Polletta} where we only use MIR-FIR part of the template for the fitting because we cropped at rest-frame wavelengths below 4.5 $\micron$ as mentioned above.
The best fit AGN template is a template with $\theta$ = 19\arcdeg, $r_{\rm in}$ = $3\times10^{13}$ cm, $\eta$ = 7.7\%, $\tau_{V,{\rm cl}}$ = 13.5, and $\tau_{V,{\rm mid}}$ = 1000 in \cite{Siebenmorgen}.
The estimated IR luminosity is  $2.2^{+1.5}_{-1.0}$ $\times 10^{14}$ $L_{\sun}$, which is classified as an ELIRG.
We remind that {\tt SEABASs} decomposes stellar, AGN, and SF components, and calculates IR luminosity for each component, providing us the AGN contribution to the IR luminosity.
The luminosity contribution of the AGN, $L_{\rm IR}$(AGN)/$L_{\rm IR}$, is $94^{+6}_{-20}$\%, which follows the $L_{\rm IR}$(AGN)/$L_{\rm IR}$ and $L_{\rm IR}$ relation reported by \cite{Toba_17b}. 

The color excess derived from the SED fitting with stellar component considering a Calzetti dust extinction law is $E (B-V)$ = 0.45 mag that can be translated to $N_{\rm H}$ by assuming the following relation (\citealt{Ricci_a}, see also \citealt{Maiolino}),
\begin{equation}
\frac{E(B-V)}{N_{\rm H}} = 1.5 \times 10^{-23}\,\,{\rm cm^2} \,\, {\rm mag}.
\end{equation}
As a result, we found $N_{\rm H} = 3.0 \times 10^{22}$ cm$^{-2}$, suggesting that WISE1013 is a mildly dust-obscured AGN.

Since this SED fitting code does not just freely scale each SED template to fit the data because of the requirement for energy conservation as described above, the resultant SED is useful to investigate the physical properties of WISE1013.
In order to confirm a robustness of physical quantities derived by our method with {\tt SEABASs}, we also  perform the SED fitting with other SED fitting codes and check the consistency of resultant physical quantities.
Here, we utilized {\tt MAGPHYS}\footnote{\url{http://www.iap.fr/magphys/}} (Multi-wavelength Analysis of Galaxy Physical Properties; \citealt{da_Cunha_08, da_Cunha_15}) and {\tt CIGALE}\footnote{\url{https://cigale.lam.fr/}} (Code Investigating GALaxy Emission; \citealt{Burgarella,Noll}), allowing us to do a detailed SED modeling in a self-consistent framework \citep[see also][]{Ciesla,Chang_17}.
We confirmed that the resultant quantities; $L_{\rm IR}$, $E(B-V)$, $L_{\rm IR}$(AGN)/$L_{\rm IR}$, and stellar mass and star formation rate (see Section \ref{M_SFR}) are in good agreement with each other, which means that the derived values that are relevant to this work are not changed within a error regardless of the SED fitting techniques, given the limited number of photometric data.
Nevertheless, it should be noted that we still do not have data at rest-frame 10-100 $\micron$ that could correspond to the peak of AGN and SF emission, which may induce the large uncertainty of the any quantities obtained from the SED fitting (see Section \ref{Dust} and \ref{M_SFR}).
Hereafter, we should keep in mind this possible uncertainty.
      
We also note that its IR luminosity is potentially amplified by the effect of beaming and/or gravitational lensing even if the derived $L_{\rm IR}$ is reliable, as discussed in \cite{Tsai}.
It is quite difficult to rule out these possibilities quantitatively based on the current dataset.
Hence, we briefly mention about these possibilities.

Since the beaming effect is related to small scale physics and thus AGNs with highly collimated outflow (e.g.,  Blazers) tends to show a variability of flux density over a wide range of wavelengths \citep[e.g.,][]{Lister}.
In order to see if WISE1013 shows variability, we checked its variability flag ({\tt var\_flag\footnote{The quantitative definition is described in the Explanatory Supplement to the AllWISE Data Release Products (\url{http://wise2.ipac.caltech.edu/docs/release/allwise/expsup/sec5_3bvi.html})}}) in the ALLWISE catalog.
This flag is a four-character string where each character gives a measure of the probability that the source is variable in each band estimated from multi-epoch photometric data.
Each character has values of 0 to 9; the objects with higher value in a band indicates higher probability of variability at that band.
On the other hand, objects with {\tt var\_flg} = ``n'' in a band means that data are insufficient or inadequate to justify a variability at that band.
We found that  WISE1013 has {\tt var\_flg} = ``n000'', which indicates that there is no sufficient high-quality photometry for variability assessment at 3.4 $\micron$ and there is no variability at 4.6, 12, and 22 $\micron$ over a timescale of a few months. 
In addition, Since WISE1013 was undetected by Faint Images of the Radio Sky at Twenty-cm (FIRST) survey \citep{Becker} where the detection limit is about 1 mJy at 20 cm, this object is unlikely to be radio-loud AGN.
These results suggest that beaming effect of this object is expected to be small.

For the possibility of lensing, we confirmed that there are no massive foreground galaxies that can act as a lens (see Figure \ref{image}), meaning that lensing effect is also expected to be small.
Nevertheless, in order to quantify this effect, we need follow-up observations with high angular resolution and need to estimate magnification factor, which is beyond the scope of this paper and will be in a future work.

	\begin{figure}
   \centering
   \includegraphics[width=0.45\textwidth]{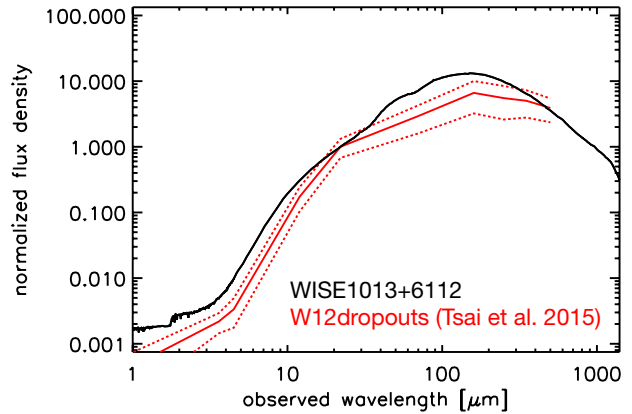}
   \caption{Comparison of best fit SED of WISE1013 (black) with a composite SED of 20 W12dropouts (red) obtained from \cite{Tsai}. Both SEDs are normalized by flux density at 22 $\micron$.}
   \label{HDOGs}
   \end{figure}
           
Recent studies have discovered many HyLIRGs and some ELIRGs based on the ``W12dropout'' method \citep{Eisenhardt,Wu}.
They are faint or undetected by WISE at 3.4 $\micron$ (W1) and 4.6 $\micron$ (W2) but are well detected at
12 $\micron$ (W3) or 22 $\micron$ (W4), and hence called ``W12dropouts'' or ``Hot DOGs''.
W12dropouts are also thought to be a key population to understand the co-evolution of galaxies and SMBHs and have been intensively investigated for their properties \citep[e.g.,][]{Assef,Fan_16a,Fan_16b}.
Figure \ref{HDOGs} shows the comparison of the best-fit SED of WISE1013 and a composite SED of 20 W12dropouts \citep{Tsai}, where both SEDs are normalized by flux density at 22 $\micron$.
Both SEDs are similar in the MIR to FIR regime, indicating that WISE1013 has similar AGN/SF properties as W12dropouts (see Section \ref{Dust}).
On the other hand, the near-IR (NIR) SED of WISE1013 is relatively bright compared to W12dropouts due to a selection effect (i.e., by definition, W12dropouts are very faint at 3.4 and 4.6 $\micron$).
Actually, WISE1013 does not satisfy the selection criteria of W12dropouts, and thus our sample selection method and W12dropout method are complementary in terms of HyLIRGs/ELIRGs search \cite[see also][]{Toba_16}.

\subsection{Dust properties of WISE1013}
\label{Dust}
    \begin{figure*}
   \centering
   \includegraphics[width=0.9\textwidth]{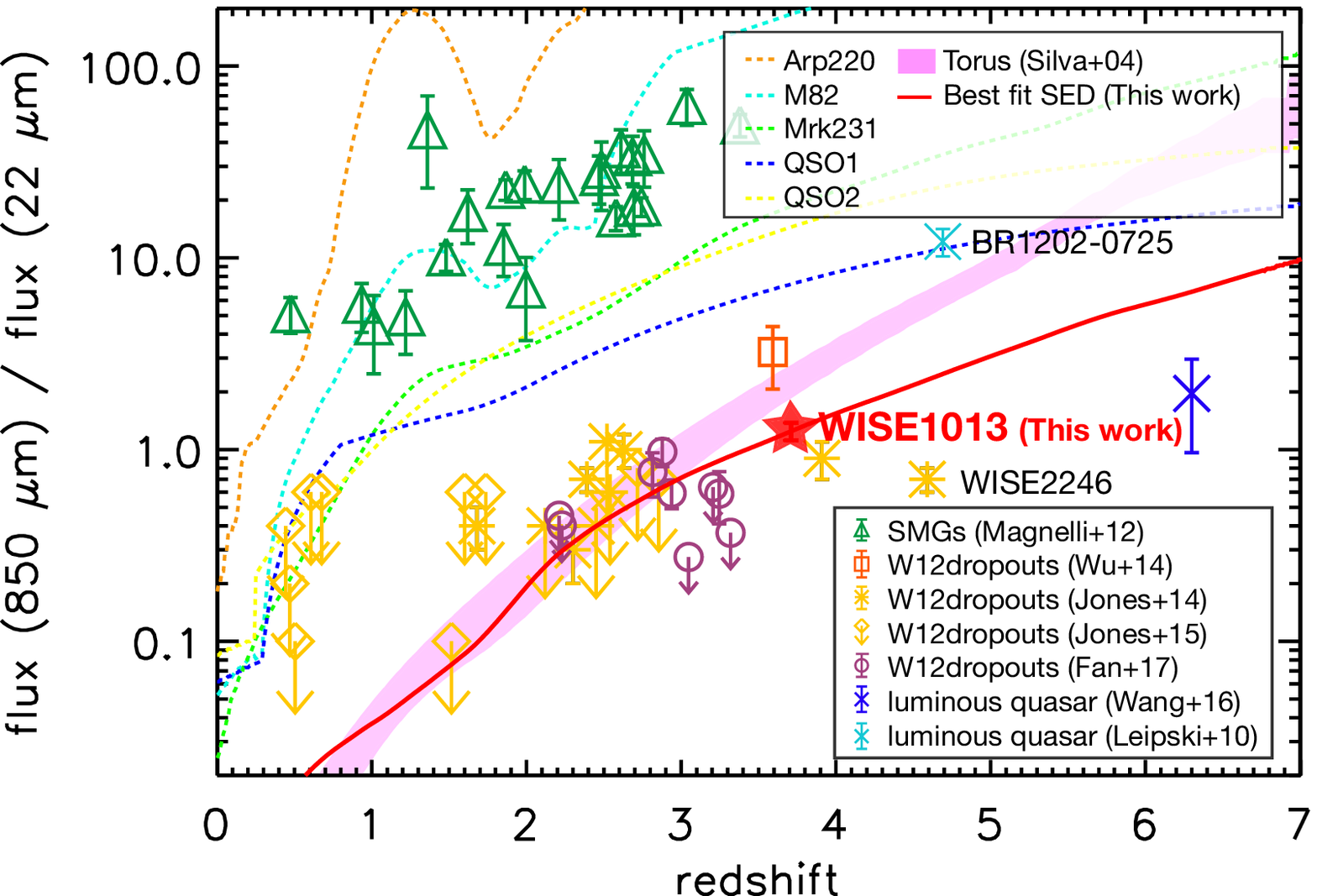}
   \caption{Ratio of 850 $\micron$ and 22 $\micron$ flux density as a function of redshift. The red star represents our sample (WISE1013). Green triangles represent SMGs \citep{Magnelli}. Orange square, yellow asterisk, yellow diamonds, and purple circles are W12dropouts obtained from \cite{Wu}, \cite{Jones_14}, \cite{Jones_15}, and \cite{Fan_17} respectively. Cyan and blue crosses represent ultraluminous quasars at $z=4.7$ \citep{Leipski} and at $z=6.3$ \citep{XWu,Wang}, respectively. Dashed lines represent flux ratios of Arp220, M82, Mrk231, type 1 quasars, and type 2 quasars calculated with SED templates \citep{Polletta}. Pink shaded region represents flux ratio estimated from torus templates with varying extinction ranging from $N_{\rm H}$ = 0 to $N_{\rm H}$ = 10$^{23}$ cm$^{-2}$ \citep{Silva}.}
   \label{ratio}
   \end{figure*}
         
Following that, we discuss the dust properties of WISE1013 and compare them with other populations.
Figure \ref{ratio} shows the ratio of flux densities at observed-frame between 850 and 22 $\micron$ (R850\_22), which trace cold and hot dust components, respectively.
This ratio could tell us which component is more dominant in a galaxy.
R850\_22 values of other populations obtained from the literature \citep{Magnelli,Wu,Jones_14,Jones_15,Wang,Fan_17} and estimated from SED templates \citep{Polletta,Silva} are also shown in this Figure.
Note that we used 24 $\micron$ flux densities instead of 22 $\micron$ flux densities for submillimeter galaxies (SMGs) and BR1202-0725 \citep{Leipski}.
We converted from 800 $\micron$ to 850 $\micron$ flux density assuming $f_{\nu} \propto \nu^{\beta+2}$ where $\beta = 1.5$ for BR1202-0725.
   
The derived R850\_22 of WISE1013 is 1.25 $\pm$ 0.13 which is significantly lower than those of SMGs, indicating that hot dust is more dominant in WISE1013 compared to the SMGs.
On the other hand, R850\_22 of WISE1013 is comparable or slightly larger than those of W12dropouts.
This indicates that the dust temperature of W12dropouts (Hot DOGs) tends to be hotter than normal DOGs, which supports the previous works \citep[e.g.,][]{Wu}.
Comparing with R850\_22 calculated using SED templates, Arp220 and M82 templates \citep{Polletta} can reproduce the R850\_22 of SMGs, while typical optically-selected AGNs cannot explain the R850\_22 of our sample.
This supports that hot dust in WISE1013 is more dominant than that in normal AGNs.  
On the other hand, most W12dropouts and WISE1013 are roughly consistent with those estimated from torus templates with varying extinction ranging from $N_{\rm H}$ = 0 to $N_{\rm H}$ = 10$^{23}$ cm$^{-2}$ \citep{Silva}.

\subsection{Star formation rate and stellar mass relation}
\label{M_SFR}

Finally, we discuss the stellar mass ($M_*$) and star formation rate (SFR) of WISE1013 at $z = 3.703$.
We estimated $M_*$ and SFR in the same manner as \citep{Toba_17b}; $M_*$ was derived from the SED fitting by {\tt SEABASs} while SFR was derived from $L_{\rm IR}$ (SF) using $\log \,{\rm SFR} = \log \, L_{\rm IR}$ (SF) - 9.966 \citep{Salim}.
The derived $M_*$ and SFR are $\log \,(M_\ast/M_{\sun})$ = $11.2^{+0.6}_{-0.2}$ and $\log \,({\rm SFR}/M_{\sun}\,{\rm yr}^{-1}$) = $3.1^{+0.2}_{-0.1}$, respectively.
Note that we confirmed that the choice of SF template with cropping at rest-frame wavelengths below 4.5 $\micron$ is insensitive to the derived SFR although {\tt SEABASs} preferred ``NGC6090'' as a best fit SF template (see Section \ref{SED_fit}). 
On the other hand, the best-fit stellar template suggests that the age of stellar population is 0.1 Gyr (Section \ref{SED_fit}) that is likely to be still young in the context of time scale of galaxy evolution; the luminosity and colors are still changing significantly with time.
We should keep in mind that the stellar mass using this stellar template could be sensitive to the choice of SED model.

   \begin{figure}
   \centering
   \includegraphics[width=0.45\textwidth]{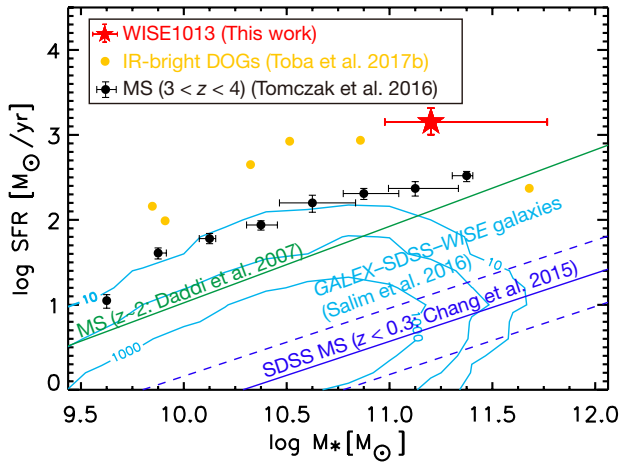}
   \caption{Stellar mass and SFR for WISE1013 (red star). The yellow circles represent those for IR-bright DOGs detected with {\it AKARI} or {\it IRAS} \citep{Toba_17b}. The blue solid line is main sequence (MS) of normal SF galaxies selected from the SDSS \citep{Chang} with scatter of 0.39 dex (blue dotted line). The cyan contours represent SFR--$M_*$ relation for a sample of GALEX--SDSS--WISE Legacy Catalog \citep[GSWLC: ][]{Salim} at $z < 0.3$. The bin size is 0.2 $\times$ 0.2 in the units given in the plot. The green line is MS of SF galaxies at $z$ = 2 \citep{Daddi}.}
   \label{M_SFR_fig}
   \end{figure}

Figure \ref{M_SFR_fig} shows the relation between stellar mass and SFR relation for WISE1013, IR-bright DOG sample detected by {\it AKARI} and/or {\it IRAS} \citep{Toba_17b}, and the main-sequence (MS)  for star forming galaxies at $3 < z < 4$ \citep{Tomczak}.
The stellar mass and SFR of the MS sample for star forming galaxies selected by the SDSS and {\it WISE} \citep{Chang}, and selected by the Galaxy Evolution Explorer \citep[{\it GALEX}:][]{Martin} satellite, SDSS, and {\it WISE} \citep{Salim} are also shown in this Figure. 
The MS presented by \cite{Daddi} for star forming galaxies at $z$ = 1 is also over plotted.
We corrected the stellar mass and SFR in the literature to those assumed Chabrier IMF if needed \citep[see][]{Toba_17b}.

We found that WISE1013 has an offset with respect to the main sequence galaxies (MS) at $3 < z < 4$ \citep{Tomczak}; given the same stellar mass, SFR of WISE1013 is about 4.2 times higher than that of star-forming galaxies at similar redshift.
This means that WISE1013 can be classified as a starburst galaxy in $M_*$-SFR plane.
Therefore, this extremely-luminous DOG shows significant AGN and SF activity that provides a good laboratory to investigate the maximum phase of galaxy--SMBH co-evolution.

\begin{figure}
   \centering
   \includegraphics[width=0.45\textwidth]{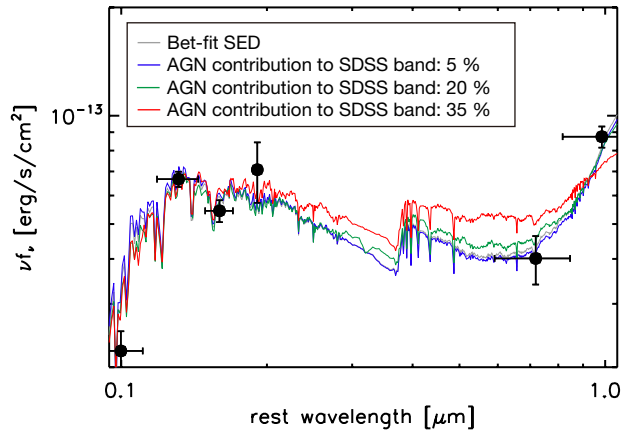}
   \caption{Influence of AGN contribution to the optical bands on the SED fitting. The black circles represent the flux at SDSS $g$-, $r$-, $i$-, $z$-band, and {\it WISE} 3.4 and 4.6 $\micron$. Gray line represents the best-fit SED without adopting any prior probability of AGN flux contribution to the optical bands (see Figure \ref{SED}). The blue, green, and red lines represent the resultant SED when assuming 5\%, 20\%, and 35\% contribution of AGN flux to the optical bands as a prior probability, respectively.} 
   \label{M}
   \end{figure}

We note that possible uncertainty of stellar mass and SFR we derived based on the SED fitting.
For the SFR, we remind that WISE1013 is AGN dominated DOG as described in Section \ref{SED_fit} and our current data set lack the rest-frame FIR data that are responsible for determining $L_{\rm IR}$ (SF).
This means that the derived SFR could have large uncertainty.
For the stellar mass, we also remind that WISE1013 has broad C{\,\sc iv} emission line with a FWHM of 5100 $\pm 160$ km s$^{-1}$ \citep[see bottom left panel of Figure 18 in][]{Hamann}.
The existence of the broad line indicates that we can see the broad line region (BLR) and radiation from AGN can contribute to the UV/optical fluxes.
Since we estimate the stellar mass based on the best-fit stellar template that fits the optical and NIR data, if AGN emission would boost optical flux, the derived stellar mass in this work could be overestimated.
Although we input type 1 AGN templates in addition to type 2/obscured AGN templates when conducting the SED fitting and most templates we input cover the optical regions, we are still not able to rule out the possibility of the AGN contribution to the derived stellar mass.
However, it is hard to estimate AGN contribution and its influence on the stellar mass precisely.
We thus discuss upper limit of the AGN contribution to the optical bands using a prior probability.
{\tt SEABASs} allows us to have prior information, e.g., the bulk of the flux in some filter comes from the AGN \citep[see][for more detail]{Rovilos}.
Figure \ref{M} shows the resultant SEDs when assuming 5\%, 20\%, and 35\% contribution of AGN flux to the SDSS bands as a prior probability.
The derived stellar mass of each case is $\log \,(M_{*}/M_{\sun})$ = 11.2, 11.3, and 11.2, respectively.
We found that {\tt SEABASs} seems to fit the data moderately well when assuming 5--20\% contribution of AGN, while it seems not to fit the data when assuming $>$ 35\% contribution as shown in Figure \ref{M}.
This suggests that the possible AGN contribution to the SDSS bands may be less than 35\%, and even if we use the best fit template for each case to derive the stellar mass, the resultant stellar mass is not significantly changed.

\section{Conclusions}
In this paper, we report the discovery of an extremely-luminous DOG (WISE J101326.25+611220.1) at $z_{\rm spec} = 3.703$.
Thanks to the multi-wavelength dataset of the SDSS, {\it WISE}, {\it AKARI}, SCUBA-2, and SMA, we pinned down its SED at rest-frame 0.1--300 $\micron$. 
We derived its physical quantities such as IR luminosity based on the SED fitting. 
Main results are summarized as follows.
\begin{enumerate}
\item The derived IR luminosity by using {\tt SEABASs} code is $L_{\rm IR}$ = $2.2^{+1.5}_{-1.0}$ $\times 10^{14}$ $L_{\sun}$, making it one of the most luminous IR galaxies in the Universe.
\item The ratio of flux densities at observed-frame between 850 and 22 $\micron$ (R850\_22) of WISE1013 is significantly lower than those of SMGs while it is comparable or slightly larger than those of W12dropouts, meaning that the dust temperature of WISE1013 is hotter than that of SMGs but it is slightly cooler than that of W12dropouts.
\item The WISE1013 covers a locus of starburst galaxies on the stellar mass and SFR plane while the AGN contribution to the IR luminosity derived from the SED fitting is about 94\%, which suggests that this extremely-luminous DOG shows significant AGN and SF activity that provides a good laboratory to investigate the maximum phase of galaxy-SMBH co-evolution.
\end{enumerate}

\acknowledgments
The authors gratefully acknowledge the anonymous referee for a careful reading of the manuscript and very helpful comments.
We are deeply thankful to Dr.Mark G. Rawlings for supporting our observations with SCUBA-2.
We also appreciate Dr.Manolis Rovilos, Dr. Elisabete da Cunha, and Dr. St\'ephanie Juneau for their helpful comments on {\tt SEABASs} and {\tt MAGPHYS}.
The Submillimeter Array is a joint project between the Smithsonian Astrophysical Observatory and the Academia Sinica Institute of Astronomy and Astrophysics and is funded by the Smithsonian Institution and the Academia Sinica.
The James Clerk Maxwell Telescope has historically been operated by the Joint Astronomy Centre on behalf of the Science and Technology Facilities Council of the United Kingdom, the National Research Council of Canada and the Netherlands Organisation for Scientific Research.
Additional funds for the construction of SCUBA-2 were provided by the Canada Foundation for Innovation. 
Funding for SDSS-III has been provided by the Alfred P. Sloan Foundation, the Participating Institutions, the National Science Foundation, and the U.S. Department of Energy Office of Science. The SDSS-III web site is http://www.sdss3.org/.
SDSS-III is managed by the Astrophysical Research Consortium for the Participating Institutions of the SDSS-III Collaboration including the University of Arizona, the Brazilian Participation Group, Brookhaven National Laboratory, Carnegie Mellon University, University of Florida, the French Participation Group, the German Participation Group, Harvard University, the Instituto de Astrofisica de Canarias, the Michigan State/Notre Dame/JINA Participation Group, Johns Hopkins University, Lawrence Berkeley National Laboratory, Max Planck Institute for Astrophysics, Max Planck Institute for Extraterrestrial Physics, New Mexico State University, New York University, Ohio State University, Pennsylvania State University, University of Portsmouth, Princeton University, the Spanish Participation Group, University of Tokyo, University of Utah, Vanderbilt University, University of Virginia, University of Washington, and Yale University.
This publication makes use of data products from the Wide-field Infrared Survey Explorer, which is a joint project of the University of California, Los Angeles, and the Jet Propulsion Laboratory/California Institute of Technology, funded by the National Aeronautics and Space Administration. 
This research is based on observations with AKARI, a JAXA project with the participation of ESA.
Y.Toba, W.H.Wang, C.F.Lim, and Y.Y.Chang acknowledge the support from the Ministry of Science and Technology of Taiwan (MOST 105-2112-M-001-029-MY3).
T.Nagao is financially supported by the Japan Society for the Promotion of Science (JSPS) KAKENHI (16H01101 and 16H03958).


\end{document}